\documentclass[runningheads,a4paper]{llncs}

\usepackage{epsfig}
\usepackage{ulem}
\usepackage{verbatim}
\usepackage{graphicx}

\usepackage{amssymb,amsfonts}
\usepackage{amsmath}
\usepackage{algorithm}
\usepackage{algpseudocode}

\usepackage{cite}

\usepackage{pifont}

\RequirePackage{ifthen}
\usepackage{amssymb,mathrsfs,wasysym}
\usepackage{tikz}
\usepackage{url}
\usetikzlibrary{shadows}
\usetikzlibrary{shapes}

\usepackage{graphicx} 

\usepackage{url}
\urldef{\mailsa}\path|jebarreno@indra.es, mvalls@it.uc3m.es|

\usepackage{authblk}

\begin{document}

\mainmatter  

\title{Supporting the monitoring of the verification process
of critical systems'software}

\titlerunning{Escribano-Barreno J. and Garc\'{i}a-Valls, M.}

\author[1,2]{Julio Escribano-Barreno}
\author[2]{Marisol Garc\'{i}a-Valls}
\affil[1]{
Indra,
Av. de Bruselas 35, 28018 Alcobendas, Spain}
\affil[2]{
Universidad Carlos III de Madrid,
Av. Universidad 30, 28911 Legan\'{e}s, Spain}


%
\authorrunning{Escribano-Barreno, J. and Garc\'{i}a-Valls, M.}

\institute{}

%
%

\toctitle{Lecture Notes in Computer Science}
\tocauthor{Authors' Instructions}
\maketitle

\begin{abstract}
Critical software systems face stringent requirements in safety, security, and reliability due to the circumstances surrounding their operation. Safety and security have progressively gained importance over the years due to the integration of hardware with software-intensive deployments that introduce additional sources of errors. It is, then, necessary to follow high-quality exhaustive software development processes that besides the needed development activities to increase safety and security also integrate techniques to increase the reliability of the software development process itself.
In practice, the use of automated techniques for the \textit{verification of the verification process} is, however, not sufficiently wide spread. This is mainly due to the high cost of the required techniques and to their degree of complexity when adjusting to the different norms and regulations.
This work presents an approach for comprehensive management of the verification processes; the approach allows engineers to monitor and control the project status regarding the applicable standards. This approach has been validated through its implementation in a tool and its application to real projects.

\end{abstract}

\section{Introduction}
\label{sec:Intro}
Failure avoidance is of paramount importance in critical software systems, 
needed to be considered in the first place as it has a direct impact on the 
safe operation. The main reasons for system failure are described in \cite{HSE2003}. Around 
44\% of the failures are caused by wrong system specifications, and a 15\% are 
introduced during the design and development phases. This means that 
over 60\% of the failures can be avoided during design and development phases. 
Some of the most common errors are to start the verification process in the 
final phase of the project or not to take into account the safety from the beginning.
On the other hand, the cost associated to certain bugs and 
reingeneering can be unacceptably high.  In general, the cost of the development 
activities is directly related to the required safety level. Although this cost tends 
to be restrained, it can be doubled if the development follows an approach that 
does not consider safety requirements.

There are a number of regulations and norms such as DO-178B\cite{DO178B} and 
DO-278\cite{DO278} that define the set of \textit{objectives} to achieve an acceptable 
level of safety. 

In the scope of critical software projects, there is a process of continuous 
investigation in order to reduce costs and development time, that usually includes 
the definition of methodologies and process automation and/or optimization techniques. 
The usage of adequate tools that support them is essential in order to optimize the 
monitoring of the processes. Using a \textit{V- Model}, the processes or phases 
identified in DO-178B\cite{DO178B} and DO-278\cite{DO278} are \textit{planning}, \textit{requirements}, \textit{design}, 
\textit{coding and integration}, \textit{integration}, \textit{verification}, \textit{configuration management}, 
\textit{quality assurance} and \textit{certification/approval liaison}. Some of these phases 
are \textit{integral processes} and must be performed through all the software life cycle.

Due to the specific requirements of the safety monitoring processes, it is usual 
that the status information for all the involved processes is very complex; information 
is typically contained in different places, and it is responsibility of several teams. 
This situation makes the status monitoring to be a difficult task to perform, prone to 
causing several undesirable effects such as missing information, or bad dimensioning of 
the development situation. 
In critical systems developments, this may lead to unsafe situations that have to be avoided.

Current monitoring tools are, in practice, complex tool chains mostly supporting separate 
activity monitoring. Usually, the different views over the development progress are provided 
by distinct tools in a non collaborative environment. Since different critical software projects 
follow different norms, there are no tools that are able to be easily customized to adapt 
to different norms. For instance, embedded software installed in aircrafts (like cockpits, 
mission computers and others) is usually developed under DO-178B\cite{DO178B} while ground 
equipments (like navigation or surveillance radars) use DO-278\cite{DO278}. In both cases,
most of the processes are similar, but they are in fact different among themselves. 
For example, DO-278\cite{DO278} introduces considerations about COTS software or adaptation data. 
Therefore, it is needed to take into account the particular characteristics of each norm, additional 
requirements introduced in each project, and possible changes in the applicable norm.

To increase the efficiency of development in critical software systems, it is required 
that norms are supported by flexible yet pragmatic and more powerful monitoring 
methodologies that can be flexible to adjust to different norms. A severe drawback 
of current methods and tools is that they do not fully support the definition and 
use of a consistent and uniform methodology for the monitoring of the verification
activities for all projects, and they do not  facilitate the compliance with regulatory 
requirements.

This paper describes an approach to improve the current practices of monitoring 
by providing a new methodology that covers the verification management activities 
in a collaborative environment to facilitate the integration with other life cycle 
process, and provide the possibility of future extensions. We present an approach 
that automates the process of monitoring by the integration of optimization mechanisms 
that includes all the information regarding the compliance statement of the 
applicable norm.

%

The paper is structured as follows. Section \ref{sec:Intro} presents an introduction and motivation for this work. Section \ref{sec:Background} describes the related work in what concerns norms and practices used for monitoring. 
Section \ref{sec:AvApproach} 
describes the proposed approach that allows cross-norm monitoring of the verification. 
Section \ref{sec:Validation} presents the implementation of the methodology in a tool and
its usage and results for a real-world critical software project. Section \ref{sec:Conclusion} concludes
the paper.

\section{Background and related work}
\label{sec:Background}

 
\subsection{Standards and engineering processes}
\label{sec:Standards_processes}

There are several standards that define processes for the software development of critical systems. Most of these standards were initially guidelines describing an approach to the regulatory requirements; they have later become de facto regulations due to their widespread adoption. Some of these standards are defined in Table \ref{Standards}.
\begin{center}
\begin{table}[h!]
\caption{Selected related standards}
\label{Standards}
\begin{tabular}{| p{1.7cm} | p{3.7cm} | p{6.6cm} |}
	\hline
	\textbf{Doc. No.} & \textbf{Title} & \textbf{Description} \\ \hline

	ESARR6 & Eurocontrol Safety Regulatory Requirement 6 – Software in ATM Functional Systems, \cite{ESARR6} & Applicable to ATM/CNS systems. Continuation of the framework defined in \cite{ESARR4} for the software. Two accepted international standards used for compliance with it are \cite{DO278} and \cite{ED153}. \\ \hline
	
	DO-178B & Software Considerations in Airborne Systems and Equipment Certification, \cite{DO178B} & One of the most accepted international standards. Used as a basis for \cite{DO278}. Recently updated to \cite{DO178C}, that includes additional objectives and it is complemented with the supplements \cite{DO330}, \cite{DO331}, \cite{DO332} and \cite{DO333}. \\ \hline
	
	DO-278 & Guidelines for Communication, Navigation, Surveillance and Air Traffic Management, \cite{DO278} & Provides guidelines for non-airborne CNS/ATM systems. It is intended to reuse the objectives included in DO-178B/ED-12B to the software contained in ATM/CNS systems, reviewing, modifying and expanding in some cases, them. This document has also been recently updated to \cite{DO278A} \\ \hline
	
	IEC 61508 & Functional safety of electrical/ electronic/ programmable electronic safety-related systems, \cite{IEC61508} & Industry automation. It is intended to be a safety standard applicable to all kinds of industry. Includes the complete safety life cycle. It has been used as a basis for other specific documents, as railway (CENELEC 50128\cite{CEN50128}), automotive industries (ISO 26262\cite{IEC61508}) or nuclear power plants (IEC 61513\cite{IEC61503}) \\ \hline
	
	CENELEC 50128 & Railway applications - Communications, signalling and processing systems, \cite{CEN50128} & Standard applicable in the railway industry. Specifies the processes and technical requirements for the development of software for programmable electronic systems for use in railway control and protection applications \\ \hline
	
	ISO 26262 & Road Vehicles - Functional Safety 
	 & Referred to the safety application in the automotive industry. Its objective is to assure the functional safety of a electric/electronic system of a motor vehicle. Developed from the \cite{IEC61508} for its specific use in the automotive industry. \\ \hline
	
	IEC 62304 & Medical device software - Software life cycle processes, Ref. \cite{IEC62304} & This norm specifies the software life cycle requirements in medical devices \\
	\hline
\end{tabular}
\end{table}
\end{center}


On the other hand, there are a lot well-known models that define the software development process. Classical approaches include Waterfall development or the V-model, that offer a complete life-cycle approach with several phases.

The Waterfall model was introduced by Winston W. Royce in 1970 (\cite{Royce}), although the term "waterfall" was used for the first time by Bell and Thayer in 1976 \cite{BellThayer}. This model defines the software development phases and the sequence among them. The end of a phase is checked through a revision that determines if the project can start the following phase or not. There is much emphasis on documentation to provide an adequate basis to further phases, improve the design and aid the accuracy of the information exchanged in the development.
%

The V-model can be considered an extension of the Waterfall model. It is not a sequence of phases moving in a linear way, but the representation of its process forms a "V". On its left side, the development phases are represented. On the right side, the verification phases can be found. The vertex of the "V" is the coding phase. 
Criticisms to it are the lack of flexibility or ineffective testing methodology applied and difficult to be strictly applied for non-trivial projects. For example, requirements are not always clearly defined by the customer, and the development cannot wait to the clarification of all of them because of schedule constraints. 
%
In projects with dynamic, non-deterministic and continuously changing requirements it is difficult to establish accurate plans in the early stages, often leading to a waste of resources and a lot of rework due to this uncertainty. The Agile methods (Agile Manifesto \cite{AgileManifesto}) appears like an adaptive, iterative and evolutionary development methodology. There are several agile software methods and process frameworks, like Scrum \cite{ScrumGuide}, Kanban \cite{Kanban}, Extreme Programming (XP) \cite{XP}, and Adaptive Software Development (ASD) \cite{ASD}, among others.

Other approaches try to balance the previous methodologies, to take advantage of their strenghts and compensate for their weaknesses, as described by Boehm and Turner in \cite{BoehmTurner2004} and \cite{BoehmTurner2003}. They support the idea of Brooks in his article \cite{Brooks}, that building software is always hard and there is not a silver bullet because of its inherent properties: complexity, conformity, changeability and invisibility. Boehm and Turner propose a risk-based approach to be used in software projects to incorporate both agile and disciplined characteristics according to the project needs.

\subsection{Missing elements for a more automated verification monitoring}


The selection of a concrete development model often comes imposed by the
company methodology, customer requirements or simply due to convenience.
This is an additional handicap in order to introduce the standard requirements 
in the development strategies, and it becomes extremely important to be able to 
address as many development models as possible with the minimum impact in 
the applicable standard.



Regardless to the methodology used, it is necessary to monitor 
and control the status of the development activities, in order to check 
if the project objectives are being achieved and to make correction in the process if needed. 

Within the scope of this work, and in order to improve the monitoring and control activity in software projects, the following goals are defined: (1) to cover the verification management activities, (2) to develop a collaborative environment, (3) to facilitate the integration with other life cycle processes and (4) to provide the possibility of future extensions and adaptation to new norms or processes.

At this point, it is important that the monitoring and control activities are as automated as possible, using the contributions of all actors involved in the project. This way, each actor needing information about the status of the project can find it almost immediately, which can contribute to reduce risks associated to the development.

Verification processes are typically achieved by the usage of
non standardized tool chains combined with ad-hoc methods, practices and
tools that constitute a competition between companies. These are typically
set at the start of a project and have to be latter heavily modified if any change
is produced or a different project is started. Therefore, it is not
easy to obtain information about these and it is even more difficult to 
get hands on specific technology.

It is necessary to look for new solutions and mechanisms that contribute to 
reach the development objectives in a more effective way, increasing the 
reliability and safety required, reducing at the same time the risk and 
costs associated to the software development. Up to the best of our knowledge,
there is no publicly available methodology (also supported by a tool) that provides a
unifying framework for the monitoring activities, i.e., for the \textit{verification of
the verification}. The presented approach overcomes this.

\section{Support for improvement of the monitoring of the verification process}
\label{sec:AvApproach}

\begin{figure}[h!]
\centering
\includegraphics[scale=0.35]{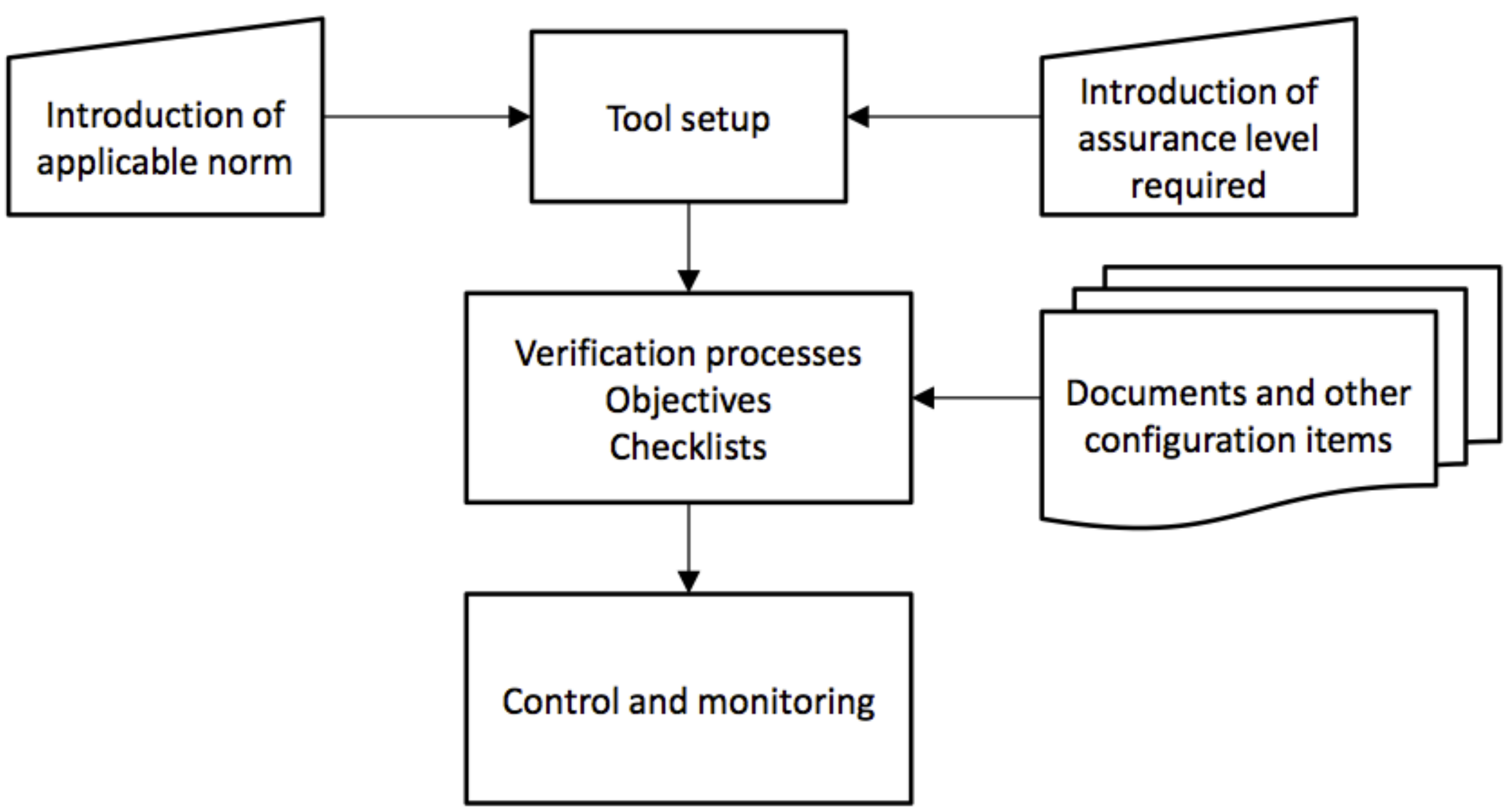}
\caption{General operational concept}
\label{tool_operational_concept}
\end{figure}

This section describes the proposed approach that aims at improving and 
automating the monitoring of the activities involved in the verification process
of critical software systems, by meeting the objectives described above. 
Figure \ref{tool_operational_concept} provides an overview of this methodology. 


Our approach relies on the initial establishment of an \textit{assurance level} and an applicable \textit{norm}. 
The \textit{assurance level} indicates the relation between a failure condition and its associated
effects in the system. \textit{Failure conditions} are classified into: \textit{catastrophic}, \textit{hazardous}, 
\textit{major}, 
\textit{minor} and \textit{no effect}. Depending on the \textit{assurance level} allocated to the software, the objectives to be reached are defined. For example, in case of DO-178B\cite{DO178B}, level A is 
the most critical level and level E is the lowest criticality one; whereas for DO-278\cite{DO278}, AL1 is the 
most restrictive one and AL6 the least restrictive one.

Derived from several years of professional engineering on monitoring of verification activities
for critical software systems in a wide range of critical software systems, our approach is based
on the following concepts: the fundamental \textit{requirements} to be met by the methodology, and a
set of operational \textit{phases}.
%

The methodology meets overall fundamental \textit{requirements}, 
that indicate the important 
objectives to be achieved in a correct verification process. The requirements 
are precisely that: ($i$) the methodology 
has to support full collaboration among all actors of the verification
process; and ($ii$) the different phases of the methodology are carried out by 
specific project roles such as developers, project managers, or verifiers  
that have different activities to perform. 
As such, it defines the access and permissions policy that determine the actions that every role is able to perform.
The methodology defines a minimum set of \textit{phases} that must be
simple in their conception in order to guarantee its applicability across projects.
The phases determine the way in which the monitoring process itself takes place, defining 
a sequence of steps to execute in order to (1) initially establish; and later (2) carry out the
monitoring of the verification activities. The methodology phases must 
allow an easy integration of different norms in the project (or across projects)
to ensure that the software complies with the specific objectives. 


%



\subsection{Specification of the monitoring framework}
\label{sec:Specification}

This initial phase defines the baseline elements of the
verification activities of the different software projects. The methodology
supports simultaneous monitoring of different projects. For a given 
software project $i$,
a number of verification activities ($VA_i$) take place to ensure the correctness 
of the software development with respect to a norm and an assurance
level. A \textit{verification activity} $j$ of a project $i$, named $V_i^j$, is an
action carried out by at least one actor; this action consist of checking a set of 
predefined elements. For each project $i$, an $E_i$ set must be
defined. It is true that for two different projects $i$ and $j$, 
it is not necessary that
the set of elements of $E_i$ and $E_j$ is equal. As the software
projects are \textit{alive}, 
the baseline elements are managed and can dynamically change over
time according to the identification of the specific project needs.

\begin{figure}[h!]
\centering
\includegraphics[scale=0.35]{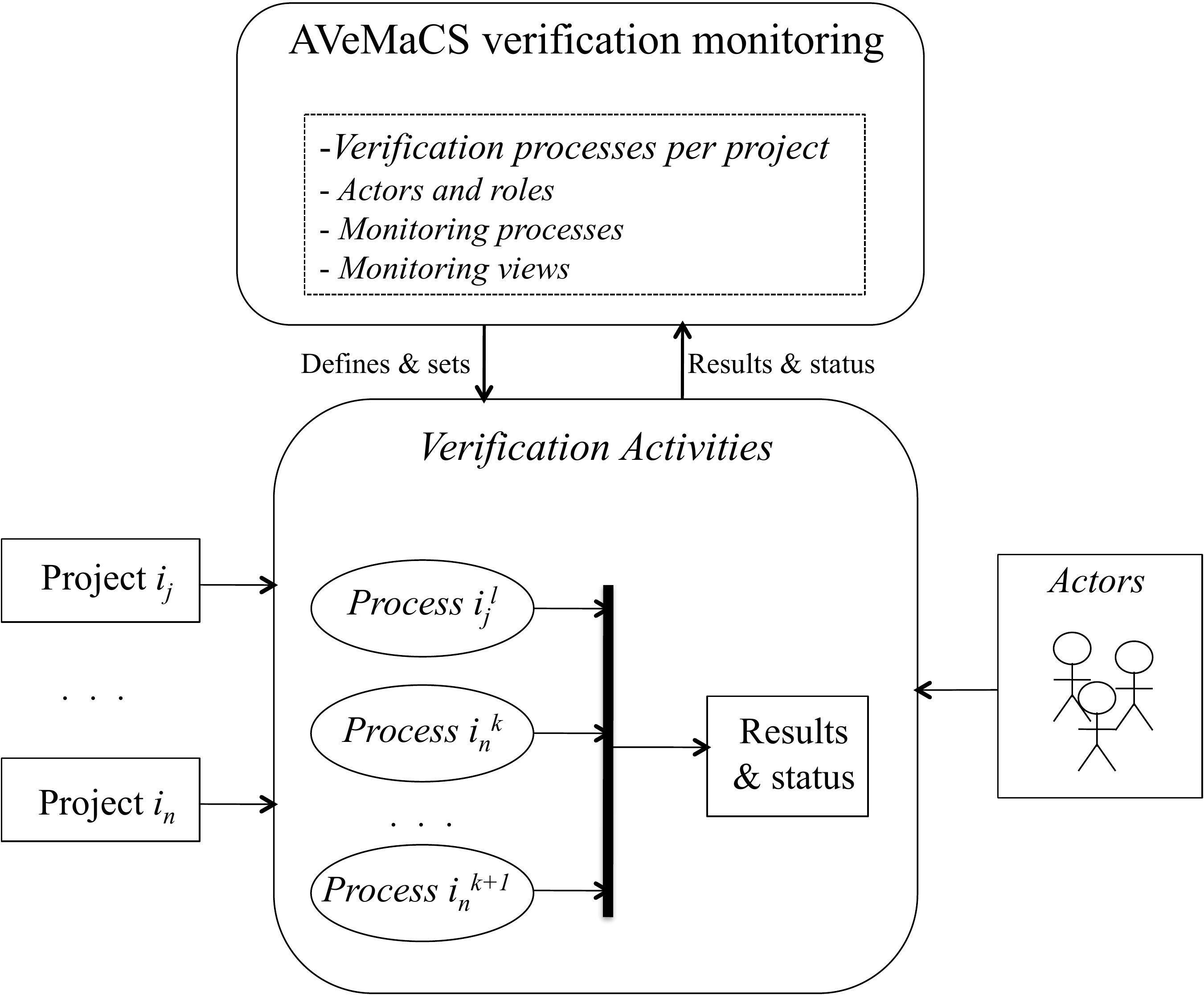}
\caption{Specification overview in the supporting tool AVeMaCS.}
\label{project_status_view}
\end{figure}


Some items must be specified
(all in sets) that define the 
operational environment of the methodology, facilitating its 
use and applicability to specific projects
and contexts. These items will be specified in section \ref{sec:parametrization}, 
as a part of the parametrization for the approach.


 

One of the essential specifications concerns the actors and their role.
Default roles are: 
administrator, verification manager, verifier, developer, and reader. 
A reader has the sole possibility of accessing the status of the development and
verification processes, and it is typically the role of project managers and technical 
managers. Developers have access to all information though they cannot modify 
the status of any item or comment. verifiers can introduce comments and 
observations about any configuration item registered for a specific process 
changing their status, and they can update the values of the verification processes 
and the configuration items and set their status. The verification manager has all 
permissions of the verification process and can update any of the elements.

\subsection{Parameterization}
\label{sec:parametrization}

This phase instantiates the above framework specifications that
are particularized to manage the verification activities. 
The progress of the verification process of a project $i$ ($prog_i$) is related
to the actual progress of the verification activities in the different processes. 
($VA_i$): $prog_i = f^{VA_i} = [Completed, Pending]$.
Progress of a project is monitored 
by defining and observing the status of the specified set of
elements $E$. When the verification process of project $i$ is completed 
then $prog_i = Completed$; value $Pending$ indicate that there is still some
comment (or \textit{non-conformity}) that has not been addressed.

From each project $i$, the following variables are extracted from 
the framework specifications of section \ref{sec:Specification}:

\begin{itemize}

\item \textit{Norm} set and standards of applicability ($NS_i$). 
\item {Assurance level} ($AL_i$). 
\item {Project characteristics} relative to specific management issues are:
\begin{itemize}
\item \textit{Life cycle} ($LC_i$) of the project. Example values (i.e., models)
          are presented in section \ref{sec:Standards_processes}.
\item The selected \textit{verification processes} ($VP_i$) to monitor the software
development correctness in its different parts. Each \textit{process} has an associated 
list of \textit{configuration items} and \textit{process checklists} that are registered for it
that reflect important information to be checked in the software development.
\end{itemize}
\item The \textit{process checklists} ($PC_{i,j}$) for each $VP_i$. If referred to 
a given project, the project subindex can be eliminated.
\item The list of \textit{configuration items} includes:
\begin{itemize}

\item Set of \textit{documents} ($DO_i$) to be provided for a given project
and norms and the status of the documents. For example, requirements or design documents.

\item The \textit{process document checklist} ($PDC_{i,j}$) for a concrete document $j$ of project
$i$ assigned to a specific verification process. If referring to a given project, it can be abbreviated eliminating the project subindex.

\item Set of \textit{observations} ($O_{i,j}$) and their status that are specific
information items of relevance.
\end{itemize}

\item \textit{Users} ($US_i$) of project $i$ including their roles. Examples of values for
user roles have been given in section  \ref{sec:Specification}.

\end{itemize}

Possible values for the above variables are later exemplified in a real project
example in Table \ref{tab:Parameterization_Example}.

%
%
%



\subsection{Assessment of verification status}

\begin{figure}[h!]
\centering
\includegraphics[scale=0.35]{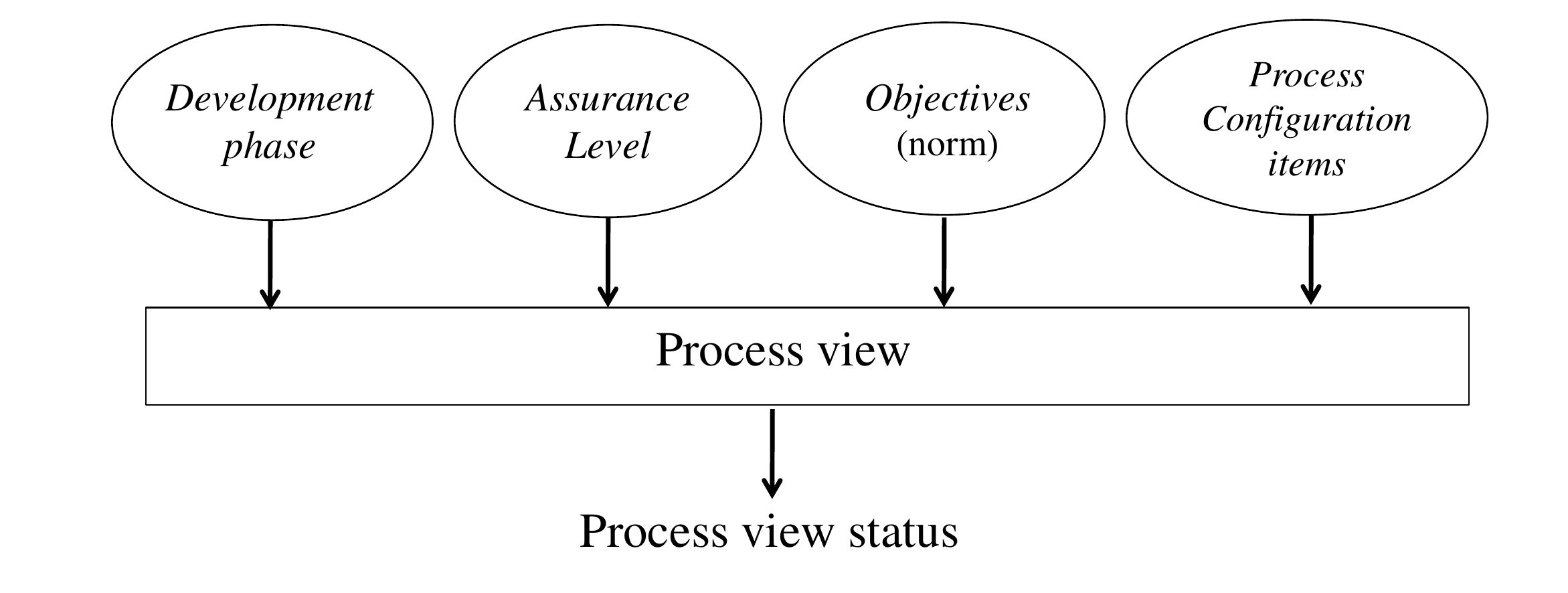}
\caption{View of the status of a given verification process}
\label{process_status_view}
\end{figure}

\begin{figure}[h!]
\centering
\includegraphics[scale=0.35]{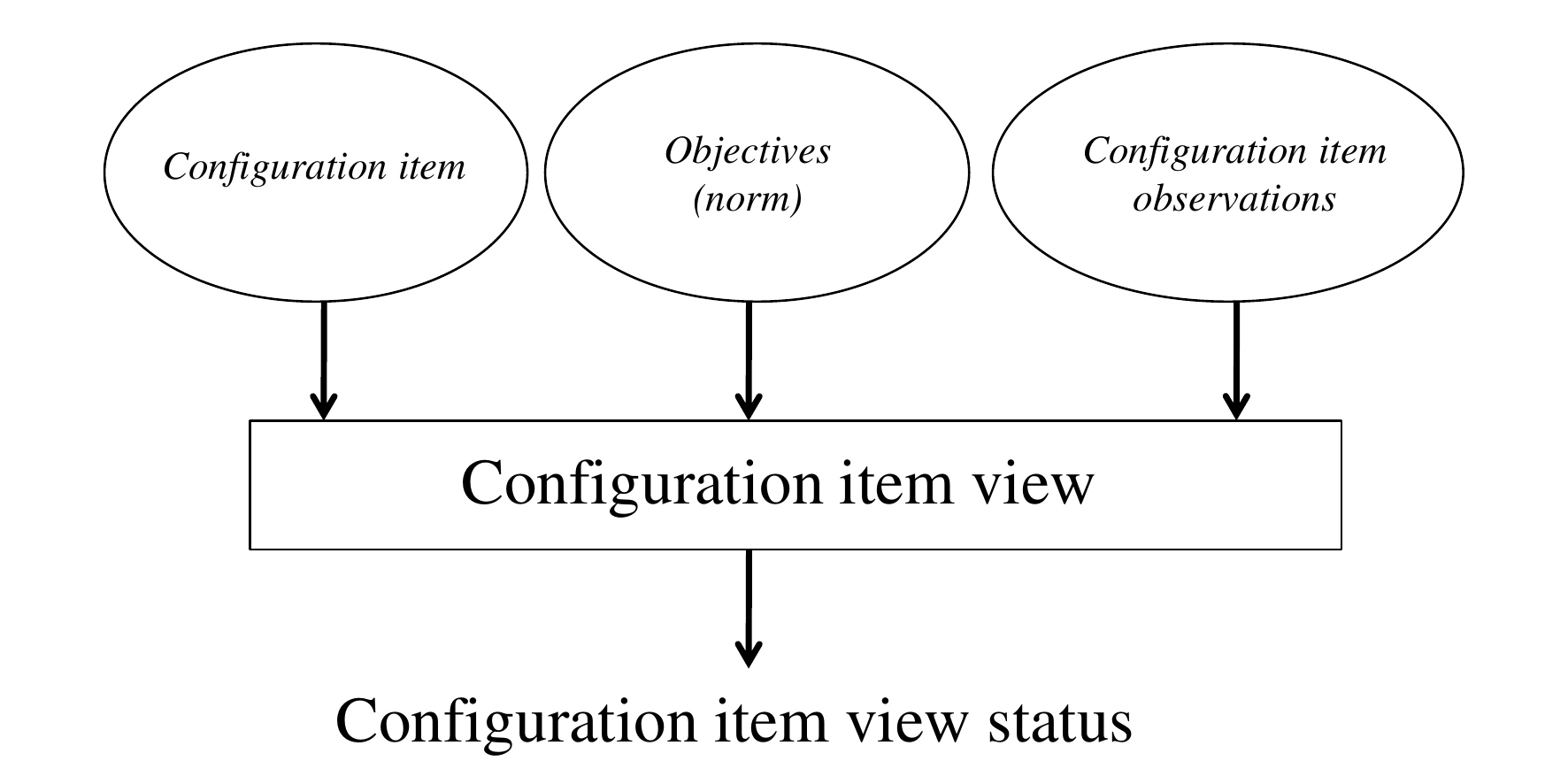}
\caption{View over the status of the configuration items}
\label{configuration_items_status_view}
\end{figure}

\begin{figure}[h!]
\centering
\includegraphics[scale=0.35]{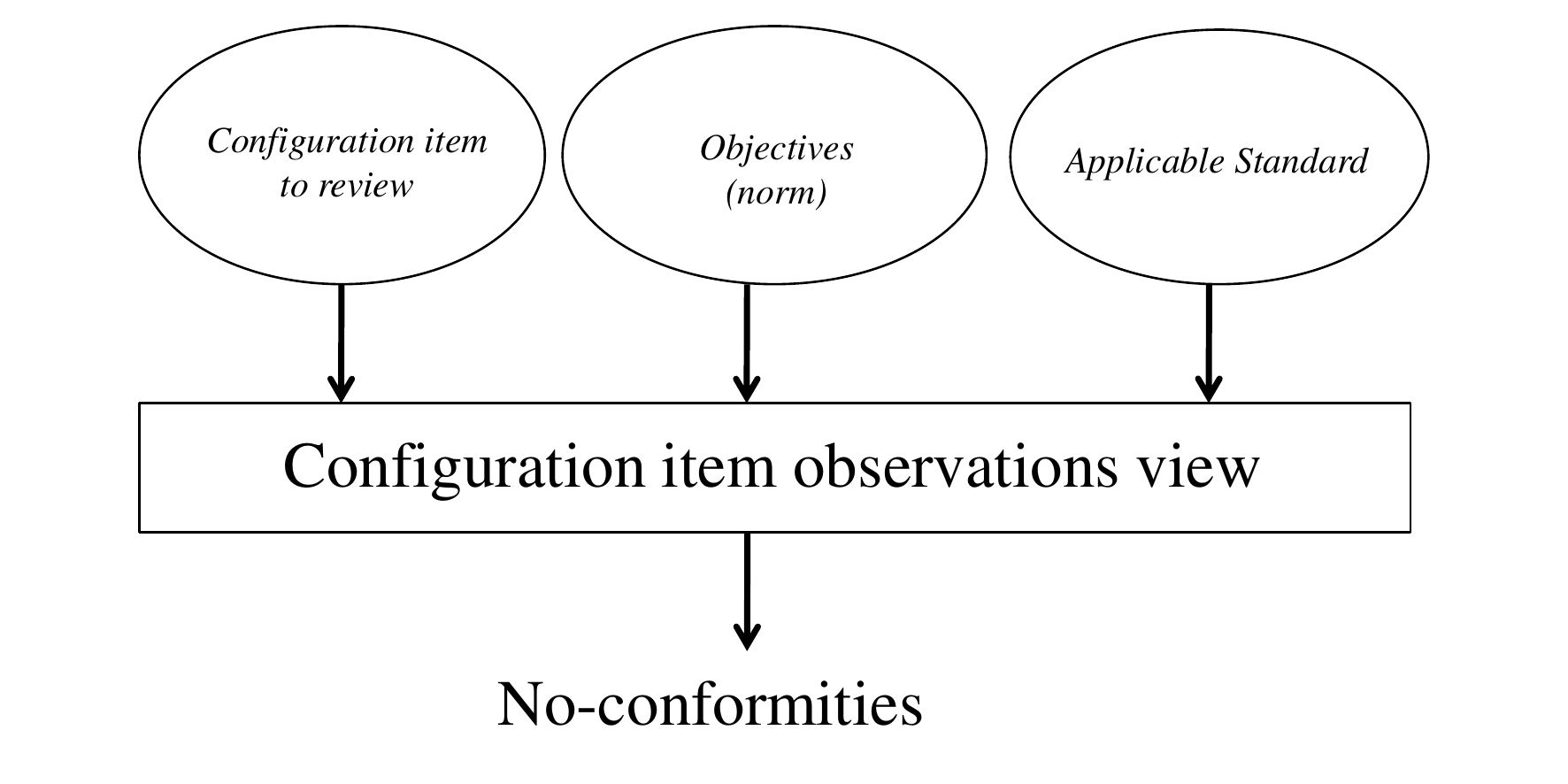}
\caption{View of the status of the configuration items' observations}
\label{configuration_items_observations_status_view}
\end{figure}

As the goal of the methodology is to assist in the 
assessment of the project status that implies the
compliance with the norm objectives, our approach
provides several levels of control or \textit{views}:

\begin{itemize}

\item \textit{Project status} view reflects the status for 
each verification process within the project. The project
status is determined by   
the \textit{Consistency and Completeness Check}, namely 
\textit{C\&C} algorithm  
(see Figure \ref{project_status_view}).

\item \textit{Verification process} view reflects the 
status of the verification processes, including the process 
status view, configuration items status view and 
the view on the configuration items observations.
\begin{itemize}

\item \textit{Process status} view reflects the results of the 
review of process objectives according to a specific norm 
and assurance level, through a process checklist (see 
Figure \ref{process_status_view}).

\item \textit{Configuration items} view shows a list of 
documents and other elements (source code, for example) 
that are determined by the used \textit{software life cycle standards}, 
including the applicable \textit{assurance standard}. It reflects the 
compliance status through the configuration items checklists and 
the observations view, described below (see 
Figure \ref{configuration_items_status_view}).

\item \textit{Configuration items observations} view shows the status of
the observations that contains the set of comments produced by 
the different actors of the verification process. 
Comments are referred to as \textit{non-conformities}.
This view has the purpose of recording the verification review, the
undertaken actions and the status, in order to provide objective 
evidences of the verification process of each item (see
Figure \ref{configuration_items_observations_status_view}).

\end{itemize}
\end{itemize}

\begin{figure}[h!]
\centering
\includegraphics[scale=0.35]{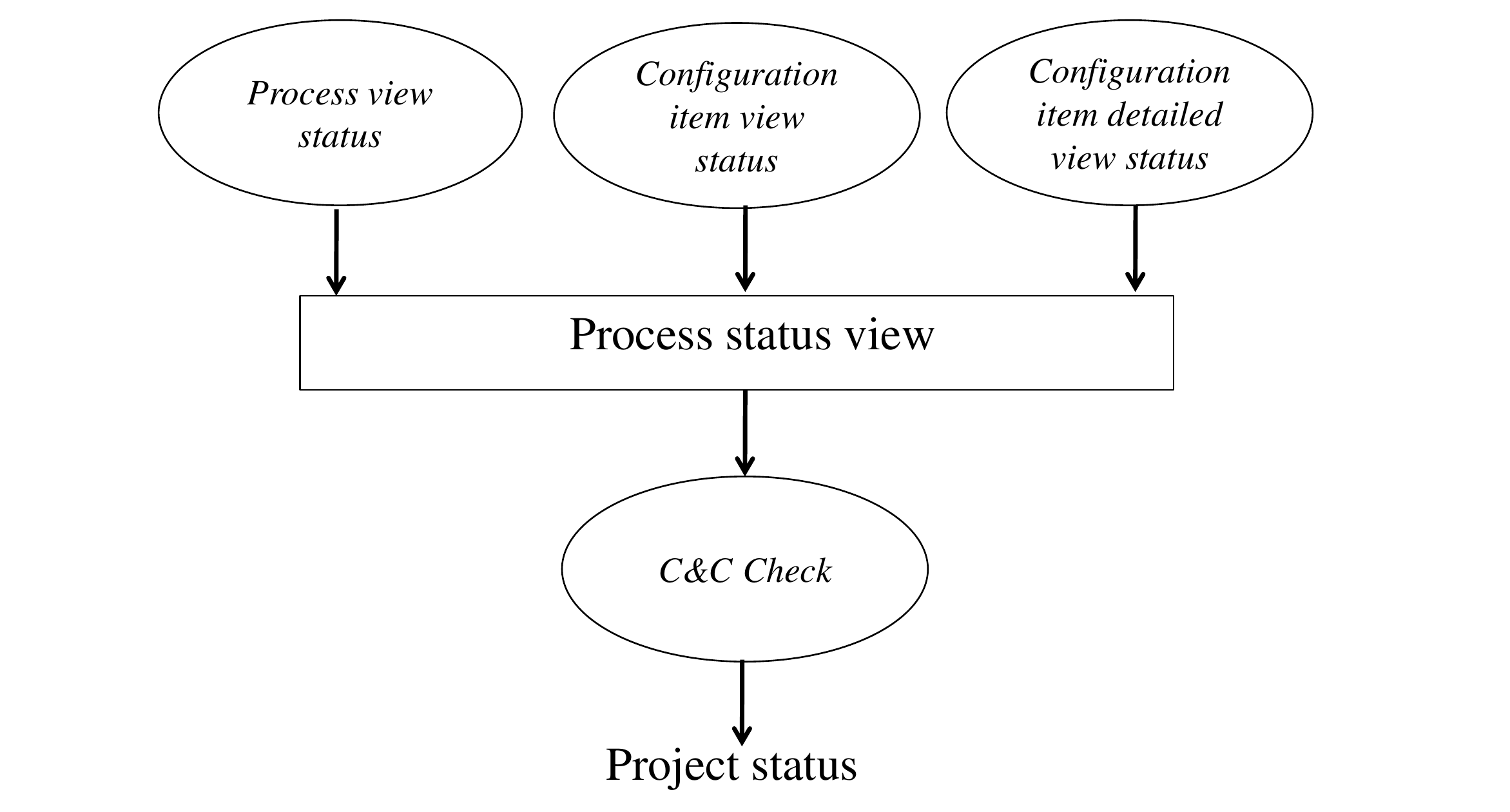}
\caption{Project status view}
\label{project_status_view}
\end{figure}
 
Each level provides information for the \textit{Consistency and 
Completeness check (C\&C check)} logic, which is shown below 
in Algorithm \ref{alg:CC}.

As an example of an objective defined by a norm that must be
checked by the verification process, we may find
\textit{"low-level requirements comply with high-level requirements"}.
This such objective is shown in the verification of outputs of Software
Design Process. If the verification actors detect that it is not met, a 
\textit{non-conformity} is opened.

The overall status of the project is determined by the partial
status of each process. In general, the process status can be either 
\textit{pending} or \textit{completed}:

\boxed{
\begin{array}{lll}
\textrm{\textbf{if} $[\forall j \; status(VP_{i,j}) = Completed]$ \textbf{then} $status(P_i) = Completed$}\\
\textrm{\textbf{else}  $status(P_{i}) = Pending$}
\end{array}}



A pending status is
assigned when there is a negative answer to any question of
the process or configuration items checklist or there is an 
opened comment for any of the configuration items.

\begin{algorithm}[h!]
\caption{C\&C Check Algorithm}
\label{alg:CC}
\begin{algorithmic}
\ForAll{($VP_j$)} 
	\State $status(PC_j)\gets Pending$
	\State $status(PDC_j)\gets Pending$
	\State $status(O_j)\gets Pending$
	\If{$PC_j$ is filled}
		\State $AnswerResults_j\gets GetProcessAnswerResults_j$ 
		\If{$AnswerResults_j == OK $}
			\State $status(PC_j)\gets Completed$
		\EndIf
	\EndIf
	\If{$PDC_j$ is filled} 
		\State $DocumentAnswerResults_j\gets GetProcessDocumentAnswerResults_j$
		\If{$DocumentAnswerResults_j == OK$}
			\State $status(PDC_j)\gets Completed$
		\EndIf
	\EndIf
	\State $status(O_j)\gets GetDocumentCommentsResults_j$
	\If{$status(O_j) == OK$}
		\State $status(PDC_j)\gets Completed$
	\EndIf
	\If{$((status(PC_j) \neq Completed)||(status(PDC_j) \neq Completed)||status(O_j) \neq Completed)$}
		\State $status(VP_j) \gets Pending$
	\Else
		\State $status(VP_j) \gets Completed$
	\EndIf	
	
\EndFor
\end{algorithmic}
\end{algorithm}

\section{Validation through a use case}
\label{sec:Validation}

The validation of the proposed approach is shown in this section firstly
by presenting its practical implementation in a software tool based on PhP 
and Java; and secondly, by presenting a real critical software project example
that has been specified and parametrized in a software tool. We show the
monitoring of the project status through different monitoring views, finally
showing the view over the observed non-conformities.

The practical use case over the tool allows to check the methodology utility
and the management of the life cycle process of the software for a given norm.

The first step was to define a norm, the objectives, and basic checklists to cover 
the whole life cycle. Once all this information is introduced in the tool, it is possible 
to modify or extend it to be reused in other projects. All the processes are 
customizable for the users (verification actors), allowing the definition of the 
scope for each project.

The project that has been selected as an example for the tool validation is
a real defence project that consists of the replacement of the cockpit display 
units of a military aircraft.
To demonstrate that the software was correctly designed, verified and validated, 
the considerations provided by \cite{DO178B} were used as an acceptable 
means of compliance. According to the Functional Hazard Analysis (see \cite{ARP4761}), 
the assurance level to be applied is DAL-B.  The development life-cycle used was a 
"V" model, and the phases defined in \cite{DO178B} were used. For each of these phases, 
the our approach was used. For the \textit{planning process}, five plans (PSAC, SDP, SVP, 
SCMP and SQAP) and three \textit{standards} (requirements, desing and coding) were 
developed, as required by \cite{DO178B}.  

\begin{center}
\begin{table}[h!]
\centering
\caption{Project parameterization}
\label{tab:Parameterization_Example}
\begin{tabular}{| p{1cm} | p{10cm} |}
	\hline
	$NS$ & DO-178B\cite{DO178B} \\ \hline
	$AL$ & DAL-B \\ \hline
	$LC$ & V-Model \\ \hline
	$VP$ & Planning, requirements, design, coding\&integration, integration, Verification of Verification \\ \hline
	$DO$ & Defined according to \cite{DO178B} \\ \hline
	$PC$ & Defined to cover \cite{DO178B} for each verification process \\ \hline
	$PDC$ & Defined to cover \cite{DO178B} for each DO \\ \hline
	$US$ & Defined according project organization and roles \\ \hline
\end{tabular}
\end{table}
\end{center}

\begin{figure}[h!]
\centering
\includegraphics[scale=0.40]{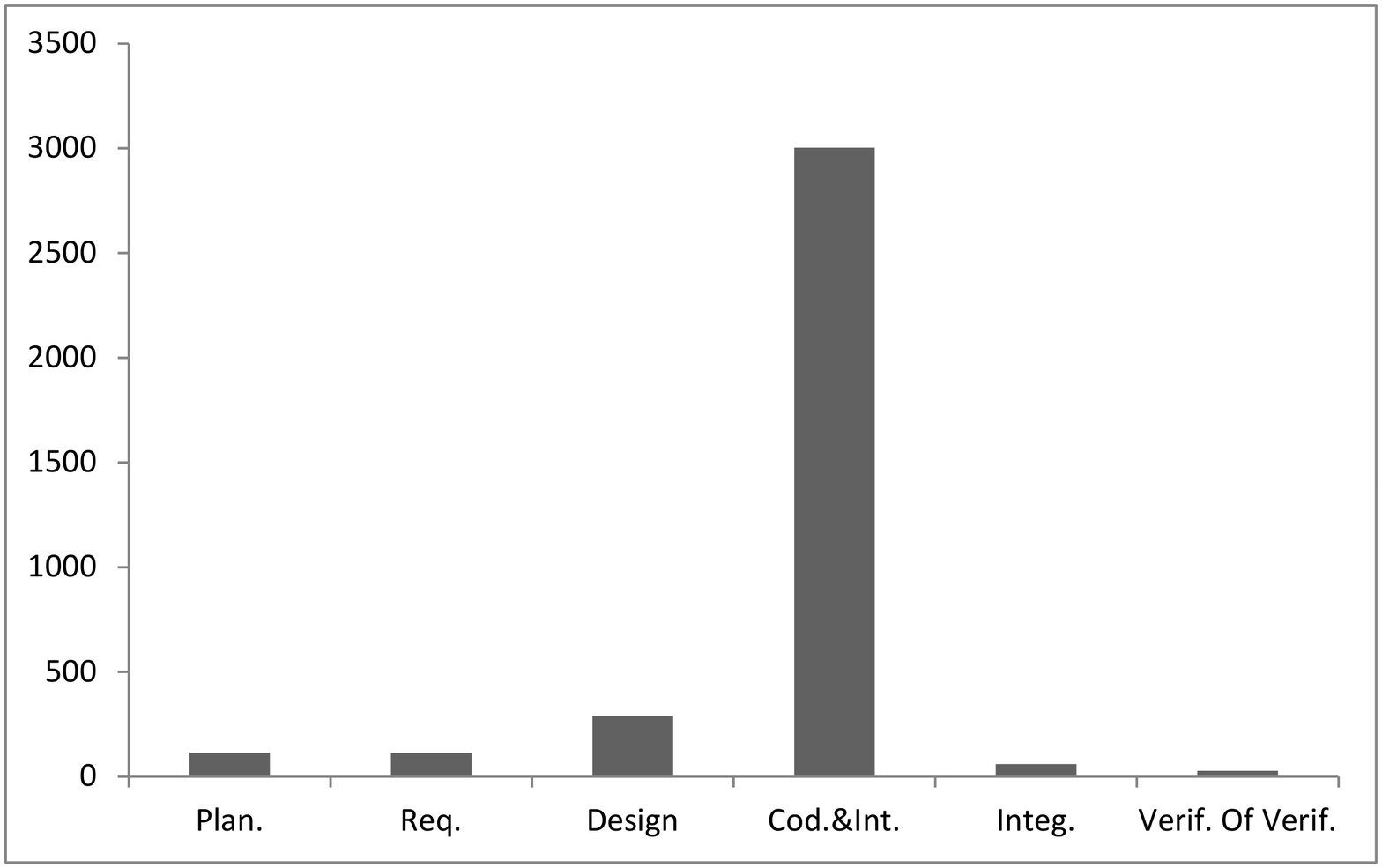}
\caption{No conformities found by process}
\label{fig:No_conformities_found_by_process}
\end{figure}

Once the project information was specified and parametrized according to Table \ref{tab:Parameterization_Example}, the monitoring process
began where all actors used the supporting tool for the monitoring of all 
verification activities. Checklists for each \textit{configuration item} were filled, showing their status and details. \textit{Observations} about each item were registered in the tool. 
Results are shown in Figure \ref{fig:No_conformities_found_by_process} where different  
metrics about the \textit{non-conformities} are found, separated by verification process.
There were $113$ observations introduced regarding the plans and standards developed 
for the project; $112$ non-conformities were registered about the requirements, and $290$ 
regarding the design; $3003$ non-conformities were registered in the coding and integration 
processes, related to the compliance with the coding standard. $60$ incidences were found 
during the execution of the tests, and were registered in the supporting tool for monitoring and 
representing the status of the integration. Eventually, $28$ non-conformities were detected 
in the \textit{verification of verification processes}. Changes needed to the procedures where 
detected during the life cycle, and introduced in the tool, as part of the verification of 
verification process.

%

The main \textit{observations} were focused on the need of including additional 
clarifications about some contents, like references to the development and 
verification tools, and the need of their qualification. Other comments refereed to 
the need of including additional information, or clarify the specified organization. 
In some cases, changes were needed such as
in the \textit{software configuration management plan} to address some processes
required by the internal company policies. The observations of the 
\textit{certification authority} were also introduced, becoming part of the information 
about the development process.

Regarding the development standards (requirements, design, and coding), there were 
not many observations given that standards are typically well known and applied in
different projects. Still additional comments were brought due to the particularities 
of each project and improvements to be introduced.

For the requirements and design processes, the requirements document was 
introduced, filling the correspondent checklists and registering the observations. 
The requirements document is one of the most important one as 
requirements are used as the basis for (1) the design as they describe the required
functionality; and (2) to define the tests to be performed in the corresponding phase. 
Consequently, there were more than $100$ of observations that involved 
changes in the original document, including some clarifications 
introduced by the author. In the review of this document, the applicable standard 
(Ref. \cite{DO178B}) and the requirements standard defined for the project were used.


%
%
%
%


The tool provided tangible evidences for the results of the verification process for 
each development process which serves as a basic reference for generate the 
\textit{Software Accomplishment Summary} and \textit{System Safety Assessment}. In these documents,
 it is necessary to include references to all the evidences needed for the development 
 assurance level required. The data introduced in the tool are able to provide an 
 evidence to the Certification Authorities of the performed work and compliance 
 status.

\subsection{Additional considerations critical software systems}

The architecture of the software is of paramount importance in critical systems
as it directly impacts the complexity of the final development and, therefore,
its verification and testing. Critical software systems verification focuses heavily
on temporal behavior applying real-time mechanisms 
(\cite{ArtistPL, ArtistQoS, ArtistAdapt,Duenas00, AresWS}). Other soft real-time
domains rather provide quality of service mechanisms embedded in the software
logic that accounds for mechanisms to allow dynamic execution whereas preserving
timely properties (\cite{qosrmFGCS, Otero, holaqos, ModeChangeICESS, Cano2013}). Verification of the properties
of distributed software also related to newer domains as cloud (\cite{rtcloudJSA}), 
the characteristics of the middleware are integrated in the model (\cite{ilandTII, Omacy}).
Lastly, specific verification mechanisms are executed on-line in very specific contexts
such as cyber-physical systems (\cite{HASE2016, Compsac2014}).

\section{Conclusions}
\label{sec:Conclusion}

The lack of information and the complexity of the applicable norms and processes in systems 
with safety requirements increases costs and difficulties to achieve  requirements compliance. 
This paper has presented an approach that clarifies these processes, that is
supported by a tool that facilitates the compliance with the safety requirements and the
adoption of new regulations.

Due to the introduced information for the data items for each development phase, 
actors involved in the projects can know the software life cycle defined with a quick view of the tool. 
This directly implies a reduction in the training time of engineers. For each item it is possible 
to know the existing non conformities, which enables their early detection and correction by
the responsible. In the tool, the collaboration is achieved through a web interface.




All information about the status of development and verification tasks is immediately
accessible to actors, that provides knowledge about the work performed and the 
pending tasks. So, an estimation of the remaining activities can be calculated, which  
could lead to implement corrective actions and minimize the impact in project goals,
decreasing the project risks.

The tool covers all the verification process according to the different levels. The supporting
tool has managed two standards (\cite{DO178B} and \cite{DO278}), covering a large scope of 
safety related projects. A set of documents and checklists for each develop phase 
and for document have been created, providing a basic framework to manage this kind 
of projects. It is possible to add additional documents at any time.


Finally, the methodology allows to follow any kind of development model. Thus, the objective 
about integration with other life cycle processes is fulfilled by the presented approach, including information 
about the development. The supporting tool provides possibilities of future growth, by 
extensions that include information about new norms or processes.

%

\bibliographystyle{IEEEtran}

\end{document}